\documentclass[twocolumn, pra, showpacs,superscriptaddress]{revtex4}
\usepackage{graphicx}
\usepackage{amsmath}
\usepackage{bm}
\usepackage{float}
\usepackage{color}

\begin{document}
\title{Quantum states of a binary mixture of spinor Bose-Einstein condensates}
\author{Z. F. Xu}
\affiliation{Department of Physics, Tsinghua University, Beijing
100084, People's Republic of China}
\author{Jie Zhang}
\affiliation{Institute of Theoretical Physics, Shanxi University,
Taiyuan 030006, People's Republic of China}
\author{Yunbo Zhang}
\affiliation{Institute of Theoretical Physics, Shanxi University,
Taiyuan 030006, People's Republic of China}
\author{L. You}
\affiliation{Department of Physics, Tsinghua University, Beijing
100084, People's Republic of China}

\date{\today}

\begin{abstract}
We study the structure of quantum states for a binary mixture of
spin-1 atomic Bose-Einstein condensates. In contrast to collision
between identical bosons, the s-wave scattering channel between
inter-species does not conform to a fixed symmetry. The
spin-dependent Hamiltonian thus contains non-commuting terms, making
the exact eigenstates more challenging to obtain because they now
depend more generally on both the intra- and inter-species
interactions. We discuss two limiting cases, where the
spin-dependent Hamiltonian reduces respectively to sums of commuting
operators. All eigenstates can then be directly constructed, and
they are independent of the detailed interaction parameters.
\end{abstract}



\maketitle

The existence of spin degree of freedom is fundamental for
elementary particles. Its values determine the
quantum statistics of atoms as bosons or fermions.
In the first experimental realization of a dilute
weakly interacting atomic Bose-Einstein condensate (BEC),
this intrinsic degree of freedom is
frozen by the external magnetic (B-) field used to spatially
confine atoms. Optical traps, on the other hand, are capable of
equal confinement of all spin components.
Remarkable experimental progresses in recent years have stimulated
many studies of atomic spinor condensates
\cite{kurn98,stenger98,barret01,griesmaier05}.

The simplest example of a spinor condensate is the spin-1 condensate
\cite{ho98,ohmi98,law98,koashi00,ho00}, whose properties are
determined by the two symmetric spin-dependent s-wave scattering
lengths. Two widely used atomic species are $^{87}$Rb and $^{23}$Na
atoms, both dominated by density dependent interactions
 in comparison to the spin exchange interactions. This
prompts the single spatial mode approximation
(SMA) for all spin components. The ground spin state is thus
solely determined by the sign of spin exchange interaction
parameter: being ferromagnetic ($^{87}$Rb) and polar ($^{23}$Na)
respectively for negative and positive spin exchange interactions.
The full quantum calculations further reveal
paired spin singlets \cite{law98,koashi00,ho00,ho00b},
different from the mean-field (MF) picture. For spin-2 condensates
\cite{koashi00,ciobanu00,ueda02}, three different s-wave
scattering channels exist, giving rise to
two spin-dependent parameters. The ground state can take the cyclic
phase in addition to the ferromagnetic and polar phases. Spin
singlets remain possible although they
can now be formed by either two or three atoms
\cite{koashi00,ueda02,ho00,ho00b}. The spin ground states become
more complicated for condensates of atoms with larger spins,
such as the spin-3 $^{52}$Cr atoms \cite{griesmaier05}.

A spinor condensate can display
remarkable non-MF features \cite{law98,koashi00,ho00,mueller06},
such as the anomalous quantum fluctuations of different
spin components and their quantum entanglement.
These strongly correlated states are also observed for
quasi-spin 1/2 bosonic gases
\cite{hall98,thalhammer08,papp08}, realized
by two distinguishable atomic species,
or two internal states of a single species
conditional on the conservations of numbers of atoms
in each spin component \cite{kuklov02,ashhab03}.

This work concerns ongoing experimental
studies of mixtures of spin-1 condensates. It
combines the features of two-component bose gases and spin-1 condensates.
Previously, the ground state phase diagrams of this type of mixtures
have been studied by us using simulated annealing under
MF approximation and SMA,
comparisons with the quantum states from
full quantum diagonalizations found general agreement with the MF states \cite{xu09}.

In this study we hope to understand
the quantum state structures including spin fluctuations
for a mixture of two spinor condensates.
We adopt
$\hat{\Psi}_{M_F}(\mathbf{r})$ $(\hat{\Phi}_{M_F}(\mathbf{r}))$ ($M_F$=-1,0,1)
as the field operator that annihilates a boson of species one (species two)
at position $\mathbf{r}$ in spin component $M_F$.
The interaction between two distinguishable spin-1 atoms
are parameterized by the scattering lengths
${a}_{0,1,2}^{(12)}$
in the channels of total spin $F_{\rm tot}=0,1$, and $2$ respectively.
The corresponding pseudo-potential is given by
${V}_{12}(\mathbf{r}_1-\mathbf{r}_2)=({g}_0^{(12)}\mathcal{P}_0
+{g}_1^{(12)}\mathcal{P}_1+{g}_2^{(12)}\mathcal{P}_2)\delta(\mathbf{r}_1-\mathbf{r}_2)/2$ \cite{xu09,mistake},
where ${g}_{0,1,2}^{(12)}=4\pi\hbar^2{a}_{0,1,2}^{(12)}/\mu$.
 $\mu$ is their reduced mass.
$\mathcal{P}_{0,1,2}$ is the corresponding projection operator,
and $\vec{F}_1\cdot\vec{F}_2=\mathcal{P}_2-\mathcal{P}_1-2\mathcal{P}_0$.
An alternative form is
\begin{eqnarray}
  {V}_{12}(\mathbf{r}_1-\mathbf{r}_2)=\frac{1}{2}(\alpha+\beta\mathbf{F}_1\cdot\mathbf{F}_2+\gamma\mathcal{P}_0)
  \delta(\mathbf{r}_1-\mathbf{r}_2),
  \label{intmsc}
\end{eqnarray}
where $\alpha=({g}_1^{(12)}+{g}_2^{(12)})/2$,
$\beta=(-{g}_1^{(12)}+{g}_2^{(12)})/2$, and
$\gamma=(2{g}_0^{(12)}-3{g}_1^{(12)}+{g}_2^{(12)})/2$.

Using the SMA with mode functions $\psi(\mathbf{r})$ and $\phi(\mathbf{r})$
for the two atomic species, the field operators can be expanded as
\begin{eqnarray}
  \hat{\Psi}_i(\mathbf{r})=\hat{a}_i\,\psi(\mathbf{r}),\qquad
  \hat{\Phi}_i(\mathbf{r})=\hat{b}_i\,\phi(\mathbf{r}),
  \label{operator}
\end{eqnarray}
where $\hat{a}_i$ and $\hat{b}_i$ are the respective
annihilation operators for the spin component $i$.
They satisfy the usual boson
commutation relations. In the absence of B-field,
the spin-dependent Hamiltonian for the mixture model we discuss becomes
\begin{eqnarray}
  H_s&=&\frac{1}{2}C_1\beta_1\left(\hat{L}_{1}^2-2\hat{N}_1\right)
  +\frac{1}{2}C_2\beta_2\left(\hat{L}_{2}^2-2\hat{N}_2\right)\nonumber\\
  && +\frac{1}{2}C_{12}\beta\, \hat{L}_1\cdot\hat{L}_2
  +\frac{1}{6}C_{12}\gamma\hat{\Theta}_{12}^{\dag}
  \hat{\Theta}_{12},
  \label{ham}
\end{eqnarray}
where the interaction parameters $C_1$, $C_2$, and $C_{12}$ take the
same definitions as in \cite{xu09}, and the inter-species singlet
pairing operator is defined as
$\hat{\Theta}_{12}^{\dag}=\hat{a}_1^{\dag}\hat{b}_{-1}^{\dag}-\hat{a}_0^{\dag}
\hat{b}_0^{\dag}+\hat{a}_{-1}^{\dag}\hat{b}_1^{\dag}$.
The atom numbers for each species are conserved,
thus operators $\hat{N}_{1,2}$ are constants
$N^{(1,2)}$. We have defined operators
$\hat{L}_{1-}\equiv\sqrt{2}(\hat{a}^{\dag}_1\hat{a}_0+\hat{a}^{\dag}_0\hat{a}_{-1})$,
$\hat{L}_{1+}\equiv\sqrt{2}(\hat{a}^{\dag}_0\hat{a}_1+\hat{a}^{\dag}_{-1}\hat{a}_0)$,
and
$\hat{L}_{1z}\equiv(\hat{a}^{\dag}_1\hat{a}_1-\hat{a}^{\dag}_{-1}\hat{a}_{-1})$;
$\hat{L}_{2-}\equiv\sqrt{2}(\hat{b}^{\dag}_1\hat{b}_0+\hat{b}^{\dag}_0\hat{b}_{-1})$,
$\hat{L}_{2+}\equiv\sqrt{2}(\hat{b}^{\dag}_0\hat{b}_1+\hat{b}^{\dag}_{-1}\hat{b}_0)$,
and
$\hat{L}_{2z}\equiv(\hat{b}^{\dag}_1\hat{b}_1-\hat{b}^{\dag}_{-1}\hat{b}_{-1})$,
which follow angular momentum algebra \cite{law98,wu96}.
Thus $\hat{L}^2_{i}$ and $\hat{L}_{iz}$ have a
complete set of common eigenvectors $|l_i,l_{iz}\rangle$.

The first three terms commute with each
other, while all of them are non-commuting with the fourth term
in (\ref{ham}). This implies the eigenstates will
depend on interaction parameters,
different from single spinor condensate. We will
consider two limiting cases; First with $\gamma=0$ as was
studied before in the Ref. \cite{luo07} and secondly when
$C_1\beta_1=C_2\beta_2=C_{12}\beta/2$. Both corresponds to
situations where the spin-dependent Hamiltonian reduces to sums of
respectively commuting operators, thus allowing for analytical
derivations of all eigenstates.

The case of $\gamma=0$ ignores the inter-species singlet pairing
interaction. The eigenstates for our model Hamiltonian (\ref{ham})
are then simply given by the simultaneous eigenstates of the four operators
$\hat{L}_1^2$, $\hat{L}_2^2$, $\hat{L}^2=(\hat{L}_1+\hat{L}_2)^2$ and
$\hat{L}_{z}=\hat{L}_{1z}+\hat{L}_{2z}$ denoted by $|l_1,l_2,l,l_z\rangle$.
Even with a nonzero B-field, including linear Zeeman shifts, the exact
eigenstates for a spin-1 condensate are known \cite{koashi00} as
$|l_1,l_{1z}\rangle_1=Z_1^{-1/2}\left(\hat{L}_{1-}\right)^{l_1-l_{1z}}
\left(a^{\dag}_1\right)^{l_1}\left(\hat{A}_0^{(2)\dag}\right)^{q_1}|\rm
vac\rangle$,
$|l_2,l_{2z}\rangle_2=Z_2^{-1/2}\left(\hat{L}_{2-}\right)^{l_2-l_{2z}}
\left(b^{\dag}_1\right)^{l_2}\left(\hat{B}_0^{(2)\dag}\right)^{q_2}|\rm
vac\rangle$, where
$\hat{A}_0^{(2)\dag}=\hat{a}_0^{\dag2}-2\hat{a}^{\dag}_1\hat{a}^{\dag}_{-1}$,
$\hat{B}_0^{(2)\dag}=\hat{b}_0^{\dag2}-2\hat{b}^{\dag}_1\hat{b}^{\dag}_{-1}$,
$Z_1$ and $Z_2$ are normalization coefficients. The total numbers
of atoms in each species are now constrained by
$N^{(j)}=l_j+2q_j$ for $j=1,2$. A complete set of basis for
the mixture of two spin-1 condensates can be constructed
from the product state
$|l_1,l_{1z}\rangle_1\otimes|l_2,l_{2z}\rangle_2$.
This corresponds to coupling of two angular momentums and we find
$|l_1,l_2,l,l_z\rangle=\sum\limits_{m_1,m_2}
\mathcal{C}_{l_1,l_{1z};l_2,l_{2z}}^{l,l_z}|l_1,l_{1z}\rangle_1\otimes|l_2,l_{2z}\rangle_2$
where $\mathcal{C}_{l_1,l_{1z};l_2,l_{2z}}^{l,l_z}$ denotes the
Clebsch-Gordan coefficient
\begin{eqnarray}
  &&\langle l_1,l_{1z};l_2,l_{2z}|l,l_z\rangle \nonumber\\
  &=&(-1)^{l_1-l_2+l_z}\sqrt{2l+1}
  \left(
  \begin{array}{ccc}
    l_1 & l_2 & l \\
    l_{1z} & l_{2z} & -l_z \\
  \end{array}
  \right).
  \label{cgcoefficient}
\end{eqnarray}
The last factor with the parentheses is the familiar Wigner
3j-symbol. The corresponding eigenvalue is given by
$E=\frac{1}{2}C_1\beta_1\left[l_1(l_1+1)-2N^{(1)}\right]
+\frac{1}{2}C_2\beta_2\left[l_2(l_2+1)-2N^{(2)}\right]
+\frac{1}{4}C_{12}\beta\left[l(l+1)-l_1(l_1+1)-l_2(l_2+1)\right]$,
under the constrains of $0\le l_1\le N^{(1)}$, $0\le l_2\le
N^{(2)}$, and $|l_1-l_2|\le l\le l_1+l_2$. The ground state is
easily found and its quantum fluctuations studied by varying the
intra- and interspecies spin-exchange interactions as well as the
linear Zeeman shift. These and other relevant details are not the
main focus of this study and will be published elsewhere
\cite{zhang09}.

For large anti-ferromagnetic
spin-exchange interaction between the two species, the MF
ground state is termed the AA phase corresponding to fully polarized
spins of each species along opposite directions, irrespective of
the natures of intra-species spin exchange interactions \cite{xu09}.
In the quantum treatment,
the above MF ground state becomes
 $Z^{-1/2}a_1^{\dag N^{(1)}}b_{-1}^{N^{(2)}}|\rm vac\rangle$.
It fails as it is not an eigensate
of the Hamiltonian. With the help of
$2\hat{L}_1\cdot\hat{L}_2=(\hat{L}_1+\hat{L}_2)^2-\hat{L}_1^2-\hat{L}_2^2$,
the ground state is found to take the form
$|N^{(1)},N^{(2)},|N^{(1)}-N^{(2)}|,l_z\rangle$
with $-|N^{(1)}-N^{(2)}|\le l_z \le |N^{(1)}-N^{(2)}|$
when $\gamma=0$ in the AA phase.
For the special case of equal populations in two species with
$N^{(1)}=N^{(2)}=N$, it reduces to the strongly correlated form
\begin{eqnarray}
  \psi^{00}_{\rm AA}=\frac{1}{\sqrt{2N+1}}\sum\limits_{m=-N}^{N}(-1)^{N-m}
  |N,m\rangle_1\otimes|N,-m\rangle_2,
  \label{aaphase}
\end{eqnarray}
which is maximally entangled as was discovered before for
a spin-1 condensate in a double well \cite{jack05}.
It appears in our model as the ground state,
instead of a dynamically created state from adiabatically tuning
the spin-exchange interaction from antiferromagnetic to ferromagnetic through
an optically induced Feshbach resonance \cite{fedichev96,bohn97,fatemi00,theis04}.

Without loss of generality,
we consider the more general case of unequal populations of $N^{(1)}>N^{(2)}$.
The operator $\hat{L}^2$ is found to be
lower bounded by $l=N^{(1)}-N^{(2)}$ instead of the nominal
minimum eigenvalue $l(l+1)=0$.
The corresponding state changes into
\begin{eqnarray}
  \psi^{ll_z}_{\rm AA}&=&(-1)^{l+l_z}\sqrt{2l+1}\sum\limits_{l_{1z},l_{2z}}
  \left(
  \begin{array}{ccc}
    l_1 & l_2 & l \\
    l_{1z} & l_{2z} & -l_z
  \end{array}
  \right)
  \nonumber\\
  &&\times
  |l_1,l_{1z}\rangle_1\otimes|l_2,l_{2z}\rangle_2,
  \label{aaphaseun}
\end{eqnarray}
where $l_1=N^{(1)}$, $l_2=N^{(2)}$, $l=N^{(1)}-N^{(2)}$, and $l_z=l_{1z}+l_{2z}$
($-l\le l_z\le l$). The 3j-symbol is given by
\begin{eqnarray}
  \left(
  \begin{array}{ccc}
    l_1 & l_2 & l \\
    l_{1z} & l_{2z} & -l_z
  \end{array}
  \right)=(-1)^{l_1-l_{1z}}\times
  \qquad\qquad\qquad\qquad\nonumber\\
  \left[\frac{(2l_2)!\,(2l)!\,(l_1+l_{1z})!\,(l_1-l_{1z})!}
  {(2l_1+1)!\,(l_2+l_{2z})!\,(l_2-l_{2z})!\,(l+l_z)!\,(l-l_z)!}\right]^{1/2}.
  \label{wigner3jun}
\end{eqnarray}
In the absence of a B-field, states $\psi^{ll_z}_{\rm AA}$
with different $l_z$ are degenerate.

The analogous state $\psi^{ll_z}_{\rm AA}$ for two
pseudo spin-1/2 Bose-Einstein condensates was discussed before
constructed from two orbitals \cite{kuklov02} or
two atomic species \cite{shi06}.
The angular momentum like state then reduces to
\begin{eqnarray}
  |l_1,l_{1z}\rangle=\frac{\left(a_{\uparrow}^{\dag}\right)^{l_1+l_{1z}}
  \left(a_{\downarrow}^{\dag}\right)^{l_1-l_{1z}}}
  {\sqrt{\left(l_1+l_{1z}\right)!}\sqrt{\left(l_1-l_{1z}\right)!}}|\rm vac\rangle,
  \label{schwingerbosons}
\end{eqnarray}
where $a_{\uparrow}$ and $a_{\downarrow}$ are the annihilation operators
for one condensate. The state $|l_2,l_{2z}\rangle$ is constructed
similarly but with the two boson operators being $b_{\uparrow}$ and
$b_{\downarrow}$ for the other condensate. When these two states are substituted into the
state of Eq. (\ref{aaphaseun}), we arrive at a simpler form as that
in \cite{kuklov02,shi06} with
\begin{eqnarray}
  \psi_{\rm AA}^{ll_z}=Z^{-1/2}\left(a_{\uparrow}^{\dag}\right)^{l+l_z}
  \left(a_{\downarrow}^{\dag}\right)^{l-l_z}\left(a^{\dag}_{\uparrow}b_{\downarrow}^{\dag}
  -a^{\dag}_{\downarrow}b^{\dag}_{\uparrow}\right)^{2l_2},
  \label{aaphaseunsb}
\end{eqnarray}
where we assume $l_1>l_2$ and $l=l_1-l_2$. This simple yet elegant state
displays strong correlation in a form that our state $\psi_{\rm AA}^{ll_z}$
for two spin-1 condensates cannot be reduced to.

We now consider the second special case with $C_1\beta_1=C_2\beta_2=C_{12}\beta/2$
with the reduced model Hamiltonian
\begin{eqnarray}
  H_s=\frac{1}{4}C_{12}\beta\hat{L}^2
  +\frac{1}{6}C_{12}\gamma\hat{\Theta}_{12}^{\dag}\hat{\Theta}_{12}-\frac{1}{2}C_{12}\beta (\hat{N}_1+\hat{N}_2). \hskip 6pt
  \label{hamsc}
\end{eqnarray}
As in Ref. \cite{koashi00,ueda02}, we define operators
$\hat{\mathcal{S}}_-=\hat{\mathcal{S}}_+^{\dag}=\hat{\Theta}_{12}$ and
$\hat{\mathcal{S}}_z=(\hat{N}_1+\hat{N}_2)/2+3$,
which satisfy the SU(1,1) commutation relations,
\begin{eqnarray}
  [\hat{\mathcal{S}}_z,\hat{\mathcal{S}}_{\pm}]=\pm\hat{\mathcal{S}}_{\pm},
  \qquad [\hat{\mathcal{S}}_{+},\hat{\mathcal{S}}_{-}]=-2\hat{\mathcal{S}}_z.
  \label{su11}
\end{eqnarray}
The Casimir operator $\hat{\mathcal{S}}^2$ that commutes with $\hat{\mathcal{S}}_{\pm}$
and $\hat{\mathcal{S}}_z$ is given by \cite{koashi00,ueda02}
\begin{eqnarray}
  \hat{\mathcal{S}}^2\equiv-\hat{\mathcal{S}}_+\hat{\mathcal{S}}_-
  +\hat{\mathcal{S}}_z^2-\hat{\mathcal{S}}_z.
  \label{casimiroperator}
\end{eqnarray}
All terms in the Hamiltonian of Eq. (\ref{hamsc}) again are found to
commute with each other, which makes it possible to find its
eigenstates as the simultaneous eigenstates for the operators
$\hat{L}^2$ and $\hat{\mathcal{S}}_+\hat{\mathcal{S}}_-$. Firstly,
we consider simultaneous eigenstates for operators
$\hat{\mathcal{S}}^2$ and $\hat{\mathcal{S}}_z$, denoted as
$|\mathcal{S},\mathcal{S}_z\rangle$ with respective eigenvalues
$\mathcal{S}(\mathcal{S}-1)$ and $\mathcal{S}_z$. Because
$\mathcal{S}_z$ remains positive and
$\hat{\mathcal{S}}_+\hat{\mathcal{S}}_-
=\hat{\mathcal{S}}_z^2-\hat{\mathcal{S}}_z-\hat{\mathcal{S}}^2$ is
positive semidefinite, the allowed values are $\mathcal{S}=n_0/2+3\
(n_0=0,1,2,\dots)$ where the minimum of $\mathcal{S}$ is limited by
the constrain $\mathcal{S}_z\ge \mathcal{S}$ and
$\mathcal{S}_z=\mathcal{S}+n_s\ (n_s=0,1,2,\dots)$. Since
$\min(\mathcal{S}_z)=3$, $\min(\mathcal{S})=3$. The constrain from
atom number conservation $N^{(1)}+N^{(2)}=2n_s+n_0$ further limits
the values of $n_0$ and $n_s$. $n_s$ can thus be interpreted as the
number of spin-singlet pairs between two species, and $n_0$ as the
number of remaining bosons.

The model Hamiltonian form Eq. (\ref{hamsc}) for this second case
is analogous to that of a spin-2 condensate \cite{koashi00,ueda02}.
Thus we can find
its exact eigenstates and eigenvalues in a similar fashion while
noting the difference of one extra degree of freedom from the
conserved numbers of atoms for both atomic species. As the operators
$\hat{\mathcal{S}}_{\pm}$ commute with $\hat{L}^2$
and $\hat{L}_{z}$,
the eigenstates can be classified according the quantum numbers
$n_0$, $n_s$, $l$, and $l_z$. Thus they can be denoted as
$|n_0,n_s,l,l_z;\lambda\rangle$, where $\lambda=1,2,\dots,g_{n_0,l}$
is used to label the degenerate orthonormal manifold, and
\begin{eqnarray}
  E&=&\frac{1}{4}C_{12}\beta\, l(l+1)+\frac{1}{6}C_{12}\gamma\, n_s(n_s+n_0+5)\nonumber\\
  &&-\frac{1}{2}C_{12}\beta(N^{(1)}+N^{(2)}).
  \label{eigenvalues}
\end{eqnarray}
For a given set of $\{n_0,n_s,l,l_z\}$, the number of degenerate states
$g_{n_0,l}$ is independent of $n_s$ and $l_z$ as
in the spin-2 case \cite{ueda02}, except now
$n_0=n_{10}+n_{20}$, $n_{10}=N^{(1)}-n_s$,
and $n_{20}=N^{(2)}-n_s$. $g_{n_0,l}$ can be found with its
generating function defined as
\begin{eqnarray}
  G(x,y,z)\equiv\sum\limits_{n_{10}=0}^{\infty}\sum\limits_{n_{20}=0}^{\infty}\sum\limits_{l=0}^{\infty}
  g_{n_{10},n_{20},l}\,x^{n_{10}}y^{n_{20}}z^l.
  \label{gfunction}
\end{eqnarray}
Following similar procedure outlined in Ref. \cite{ueda02},
we obtain the generating function
\begin{eqnarray}
  G(x,y,z)=\frac{1+xyz}{(1-xz)(1-yz)(1-x^2)(1-y^2)}.
  \label{gfunction2}
\end{eqnarray}
Expanding $G(x,y,z)$ around $x=y=z=0$ then gives the degeneracy
factor $\lambda$ of the state $|n_0,n_s,l,l_z;\lambda\rangle$.

To construct the explicit form for the eigenstates, we compute the generating
function $G_g(x,y,z)$ for the maximum spin states $|l,l_z=l\rangle$, which is
\begin{eqnarray}
  &&G_g(x,y,z)\nonumber\\
  &&=\sum\limits_{N^{(1,2)}=0}^{\infty}\,\sum\limits_{l=0}^{\infty}
  \tilde{h}_{N^{(1)},N^{(2)},l}x^{N^{(1)}}y^{N^{(2)}}z^l\nonumber\\
  &&=\frac{1}{2\pi i}\oint_{\mathcal{L}}d\zeta\frac{(1-\zeta^{-1})}{\zeta-z}\prod_{j=-1}^1
  (1-x\zeta^j)^{-1}(1-y\zeta^j)^{-1}\nonumber\\
  &&=\frac{1+xyz}{(1-xz)(1-yz)(1-x^2)(1-y^2)(1-xy)},
  \label{ggfunction}
\end{eqnarray}
where $\tilde{h}_{N^{(1)},N^{(2)},l}$ is the total number of maximum spin states $|l,l_z=l\rangle$
for a mixture of spin-1 condensates, and the loop $\mathcal{L}$ is along the unit circle.
By expanding the generating function $G_g(x,y,z)$, we find the six
building blocks for constructing the eigenstates $|l,l_z=l\rangle$,
\begin{eqnarray}
  \hat{A}^{(1)\dag}_1&=&\hat{a}_1^{\dag},\nonumber\\
  \hat{A}^{(2)\dag}_0&=&\hat{a}_0^{\dag 2}-2\hat{a}_1^{\dag}\hat{a}_{-1}^{\dag},\nonumber\\
  \hat{B}^{(1)\dag}_1&=&\hat{b}_1^{\dag},\nonumber\\
  \hat{B}^{(2)\dag}_0&=&\hat{b}_0^{\dag 2}-2\hat{b}_1^{\dag}\hat{b}_{-1}^{\dag},\nonumber\\
  \hat{\Gamma}^{(1,1)\dag}_0&=&\hat{\Theta}_{12}^{\dag},\nonumber\\
  \hat{\Gamma}^{(1,1)\dag}_1&=&\frac{1}{\sqrt{2}}\left(\hat{a}^{\dag}_1\hat{b}^{\dag}_{0}
  -\hat{a}^{\dag}_{0}\hat{b}^{\dag}_1\right).
  \label{buildingblocks}
\end{eqnarray}
The general structure of the state $|l,l\rangle$
is then given by
\begin{eqnarray}
  &&|l,l\rangle=\sum \mathcal{C}(\{u_i\},\{v_i\},\{w_i\})
  \left(\hat{A}_1^{(1)\dag}\right)^{u_1}
  \left(\hat{A}_0^{(2)\dag}\right)^{u_2}
  \nonumber\\
  &&\times
  \left(\hat{B}_1^{(1)\dag}\right)^{v_1}
  \left(\hat{B}_0^{(2)\dag}\right)^{v_2}
  \left(\hat{\Gamma}_0^{(1,1)\dag}\right)^{w_1}
  \left(\hat{\Gamma}_1^{(1,1)\dag}\right)^{w_2}|\rm vac\rangle,\qquad
  \label{llstate}
\end{eqnarray}
where $u_i$, $v_i$, and $w_i$ satisfy the following constrains
\begin{eqnarray}
  u_1+2u_2+w_1+w_2&=&N^{(1)},\nonumber\\
  v_1+2v_2+w_1+w_2&=&N^{(2)},\nonumber\\
  u_1+v_1+w_2&=&l,
  \label{llstatecc}
\end{eqnarray}
and additionally $w_2=0,1$ from the generating function of
$G_g(x,y,z)$. For a fixed $l$, the states
\begin{eqnarray}
&&|u_1,u_2,v_1,v_2,w_1,w_2\rangle\nonumber\\
&&=Z^{-1/2}
  \left(\hat{A}_1^{(1)\dag}\right)^{u_1}
  \left(\hat{A}_0^{(2)\dag}\right)^{u_2}\left(\hat{B}_1^{(1)\dag}\right)^{v_1}\nonumber\\
  &&\quad\times\left(\hat{B}_0^{(2)\dag}\right)^{v_2}\left(\hat{\Gamma}_0^{(1,1)\dag}\right)^{w_1}
  \left(\hat{\Gamma}_1^{(1,1)\dag}\right)^{w_2}|\rm vac\rangle
  \label{states}
\end{eqnarray}
   satisfying the Eq.
(\ref{llstatecc}) forms a subspace $\mathcal{B}$. The construction
of the eigenstates $|n_0,n_s,l,l_z;\lambda\rangle$ follows Ref.
\cite{ueda02}. We consider a series of subspaces
$\mathcal{H}_{(l_z=l)}
=\mathcal{H}^{(0)}\supset\mathcal{H}^{(1)}\supset \cdots$ with
$\mathcal{H}^{j}$ spanned by states satisfying $n_s\ge j$. A new
subspace $\mathcal{B}'$ can be constructed by projecting the basis
in the subspace $\mathcal{B}$ of definite values for $n_s$. It can
be simply realized by
\begin{eqnarray}
  &&(\hat{P}^{(w_1)}-\hat{P}^{(w_1+1)})|u_1,u_2,v_1,v_2,w_1,w_2\rangle\nonumber\\
  &&=(\hat{P}^{(0)}-\hat{P}^{(w_1+1)})|u_1,u_2,v_1,v_2,w_1,w_2\rangle\nonumber\\
  &&\quad-(\hat{P}^{(0)}-\hat{P}^{(w_1)})|u_1,u_2,v_1,v_2,w_1,w_2\rangle\nonumber\\
  &&=\left(\hat{\Gamma}_0^{(1,1)\dag}\right)^{w_1}\hat{P}_{(n_s=0)}
  \left(\hat{A}_1^{(1)\dag}\right)^{u_1}
  \left(\hat{A}_0^{(2)\dag}\right)^{u_2}
  \left(\hat{B}_1^{(1)\dag}\right)^{v_1}
  \nonumber\\
  &&\qquad\times
  \left(\hat{B}_0^{(2)\dag}\right)^{v_2}
  \left(\hat{\Gamma}_1^{(1,1)\dag}\right)^{w_2}|\rm vac\rangle,\qquad
  \label{schmidtorthogonal2}
\end{eqnarray}
where $\hat{P}_{(n_s=0)}\equiv \hat{P}^{(0)}-\hat{P}^{(1)}$ is the
projection operator onto the subspace with $n_s=0$, and the
eigenvalue of the operator $\hat{\mathcal{S}}_+\hat{\mathcal{S}}_-$
is precisely zero. We have thus constructed simultaneous eigenstates
for the operators $\{\hat{\mathcal{S}}_+\hat{\mathcal{S}}_-,
\hat{{L}}^2,\hat{L}_z\}$ with $l_z=l$.
 The eigenstates for $l_z<l$ can be constructed by
simply applying $(\hat{L}_-)^{l-l_z}$, and are given by
\begin{eqnarray}
  &&(\hat{L}_-)^{\Delta l}\left(\hat{\Gamma}_0^{(1,1)\dag}\right)^{w_1}\hat{P}_{(n_s=0)}
  \left(\hat{A}_1^{(1)\dag}\right)^{u_1}
  \left(\hat{A}_0^{(2)\dag}\right)^{u_2}
  \nonumber\\
  &&\quad\times\left(\hat{B}_1^{(1)\dag}\right)^{v_1}
  \left(\hat{B}_0^{(2)\dag}\right)^{v_2}
  \left(\hat{\Gamma}_1^{(1,1)\dag}\right)^{w_2}|\rm vac\rangle,\qquad
  \label{eigenstates}
\end{eqnarray}
with $u_1,u_2,v_1,v_2,w_1=0,1,2,\dots,\infty$, $w_2=0,1$, and
$\Delta l=0,1,\dots,2l$ satisfy the following relations
\begin{eqnarray}
  n_0&=&u_1+2u_2+v_1+2v_2+2w_2,\nonumber\\
  n_s&=&w_1,\nonumber\\
  l&=&u_1+v_1+2w_2,\nonumber\\
  l_z&=&l-\Delta l.
  \label{relation}
\end{eqnarray}
The corresponding eigenvalues are given by Eq.
(\ref{eigenvalues}).

Before conclusion, we note that for the most general case where all
interaction parameters are free to take any values, we can use the
method of numerical diagonalization to find all eigenstates and the
ground states. For parameters near the two exactly solvable cases
considered here, perturbation theory can be adopted to find the the
approximate eigenstates and the ground states. With the generating
function for the maximum spin states $|l,l_z=l\rangle$ in hand, and
recognizing that they form a subspace $\mathcal{B}$, the
diagonalization can be carried out within each subspace
$\mathcal{B}$ to find the eigenstates for $l_z=l$. The states for
$l_z\ne l$ can be generated by applying the operator
$\hat{L}_-^{l-l_z}$ to the eigenstates of $l_z=l$. This way we
generate the complete structure of the eigenstates, despite of
that the model Hamiltonian contains non-commuting operators
and the actual eigenstates depend on the interaction parameters.

In conclusion, we have discussed two special cases of exact quantum
states for a mixture of two spin-1 atomic condensates: one without
inter-species singlet pairing interaction and the other when
$C_1\beta_1=C_2\beta_2=C_{12}\beta/2$. Both cases reduce the
spin-dependent Hamiltonian to sums of commuting operators, whose
eigenstates can be constructed by finding the simultaneous
eigenstates for all operators. For the first case with the
inter-species pairing interaction absent $\gamma=0$, we have further
compared the results to the MF approximation. In the interesting
case of the AA phase, we find the exact eigenstates corresponds to a
maximal entanglement between condensates of the two species.

This work is supported by NSF of China under Grant No. 10640420151, No. 10774095
and NKBRSF of China under Grant No. 2006CB921206, No.
2006AA06Z104, No. 2006CB921102 and No. 2010CB923103.
YZ is also supported by the NSF of Shanxi Province under grant No. 2009011002. \\[24pt]

Note added: This work has been ongoing for a while. It is urgent to
complete the manuscript now because we find a recent submission to
the archive by Yu Shi (arXiv:0912.2209) considered the same model
system although focusing on the special case of $\gamma=0$. Our
results agree especially in the form of the strongly correlated
state of Eq. (\ref{aaphase}) in the AA phase for equal populations
in the two atomic species. His major result that the state in Eq.
(\ref{aaphase}) can be further expressed in a more tight form as
$\psi_{\rm AA}^{00}=Z^{-1/2}\hat{\Gamma}_0^{(1,1)\dag N}|\rm
vac\rangle$, however, is wrong. To see this clearly, we consider the
simple case of two atoms in each species. The AA phase now takes the
form $\psi_{\rm AA}^{00}=\frac{1}{2\sqrt{5}}
\left[\hat{\Gamma}^{(1,1)\dag
2}_0-\frac{1}{3}\hat{A}^{(2)\dag}_0\hat{B}^{(2)\dag}_0\right]$, not
of the form given in arXiv:0912.2209. For the more general case,
with equal populations in the two species, we can find all possible
solutions of Eq. (\ref{llstatecc}) with $l=|N^{(1)}-N^{(2)}|=0$ and
$l_z=l=0$ which forms a subspace $\mathcal{B}$ constructed by the
allowed basis states $|u_1,u_2,v_1,v_2,w_1,w_2\rangle$. We can
compute the matrix elements of the operators $\hat{L}_1^2$ and
$\hat{L}_2^2$ within the subspace $\mathcal{B}$, and diagonalize to
find the eigenstates and eigenvalues with
$\langle\hat{L}_1^2\rangle=N^{(1)}(N^{(1)}+1)$, and
$\langle\hat{L}_2^2\rangle=N^{(2)}(N^{(2)}+1)$. The state of the AA
phase in Eq. (\ref{aaphase}) when $\gamma=0$ is then constructed by
a linear superposition of all possible basis states in the subspace
$\mathcal{B}$. With the increasing of atom numbers, more basis
states from the Eq. (\ref{llstatecc}) will be included, and the AA
phase in Eq. (\ref{aaphase}) will deviate more from the state
$Z^{-1/2}\hat{\Gamma}_0^{(1,1)\dag N}|\rm vac\rangle$.


\begin{thebibliography}{10}
\bibitem{mistake}
  We have previously missed a coefficient of $\frac{1}{2}$ in Ref. \cite{xu09}.
  The results we reported, however, are not affected because the
  interaction parameters $g_{0,1,2}^{(12)}$ were properly scaled.
\end{thebibliography}
\end{document}